\documentclass[noshowpacs,amsmath,
twocolumn,
superscriptaddress,
8pt,
aps,prb]{revtex4-1}
\usepackage{setspace}
\usepackage{amsmath}
\usepackage{graphicx}
\usepackage[nearskip,margin = 0pt]{subfig}

\usepackage{verbatim}
\usepackage{amsfonts}
\usepackage{amssymb}
\usepackage{epstopdf} 
\usepackage{xcolor}
\usepackage{verbatim}
\DeclareGraphicsExtensions{.pdf,.eps,.png,.jpg,.mps} 
\begin{document}
\title{Counter-propagating solitons in microresonators}

\author{Qi-Fan Yang$^{\ast}$, Xu Yi$^{\ast}$, Ki Youl Yang, and Kerry Vahala$^{\dagger}$\\
T. J. Watson Laboratory of Applied Physics, California Institute of Technology, Pasadena, California 91125, USA.\\
$^{\ast}$These authors contributed equally to this work.\\
$^{\dagger}$Corresponding author: vahala@caltech.edu}

\date{\today}

\maketitle


\noindent {\bf Solitons occur in many physical systems when a nonlinearity compensates wave dispersion. Their recent formation in microresonators opens a new research direction for nonlinear optical physics and provides a platform for miniaturization of spectroscopy and frequency metrology systems. These microresonator solitons orbit around a closed waveguide path and produce a repetitive output pulse stream at a rate set by the round-trip time. In this work counter-propagating solitons that simultaneously orbit in an opposing sense (clockwise/counter-clockwise) are studied. Despite sharing the same spatial mode family, their round-trip times can be precisely and independently controlled. Furthermore, a state is possible in which both the relative optical phase and relative repetition rates of the distinct soliton streams are locked. This state allows a single resonator to produce dual-soliton frequency-comb streams having different repetition rates, but with high relative coherence useful in both spectroscopy and laser ranging systems.} 

The recent demonstration of optical solitons in microresonators has opened a new chapter in nonlinear optical phenomena \cite{herr2014temporal,yi2015soliton,brasch2016photonic,wang2016intracavity,joshi2016thermally}. These dissipative solitons \cite{ankiewicz2008dissipative} use the Kerr nonlinearity to balance wave dispersion and to compensate cavity loss \cite{leo2010temporal,herr2014temporal}. The resulting dissipative Kerr solitons (DKSs) exhibit Raman-related phenomena \cite{milian2015solitons,karpov2016raman,yi2016theory,yang2016stokes}, optical Cherenkov radiation \cite{brasch2016photonic,matsko2016optical,yang2016spatial} and can form ordered arrays called soliton crystals \cite{cole2016soliton}.  Soliton mode locking also creates a new and very stable frequency microcomb with distinct advantages over earlier microcombs\cite{kippenberg2011microresonator}. For example, internal broadening of these combs by dispersive-wave generation \cite{brasch2016photonic} enables offset frequency measurement for comb self referencing\cite{brasch2017self}. Also, the soliton repetition rate has an excellent phase noise stability \cite{liang2015high,yi2015soliton,yi2016single} and its spectral envelope is stable and reproducible so that the resulting microcombs are suitable for dual comb spectroscopy \cite{suh2016microresonator,dutt2016chip,pavlov2017soliton}. 

\begin{figure}[!ht]
\captionsetup{singlelinecheck=off, justification = RaggedRight}
\includegraphics[width=8.5cm]{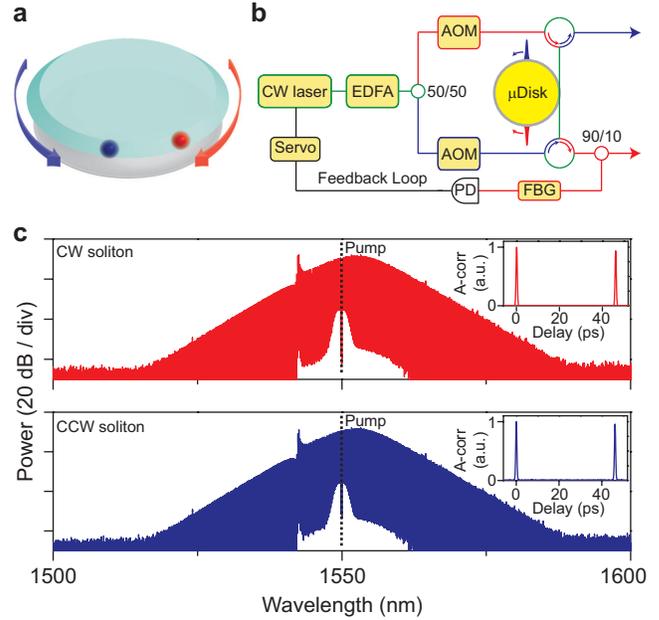}
\caption{{\bf Observation of counter-propagating solitons.} {\bf a,} Rendering showing counter-propagating solitons within a high-Q wedge resonator.  {\bf b,} Experimental setup. A continuous-wave fiber laser is amplified by an erbium-doped-fiber-amplifier (EDFA) and sent into two acousto-optic modulators (AOM). The outputs from the AOMs are counter-coupled into the microresonator and generate counter-propagating solitons. The optical power of solitons in one direction is used to Servo lock the pump laser to a certain frequency detuning relative to the cavity mode \cite{yi2016active}. FBG: fiber-Bragg grating; PD: photodetector. {\bf c,}  Optical spectra of counter-propagating   solitons. The location of the pump line is indicated by a dashed line. Measured autocorrelation traces are provided as insets.}
\label{figure1}
\end{figure}

\begin{figure*}
\captionsetup{singlelinecheck=no, justification = RaggedRight}
\includegraphics[width=18cm]{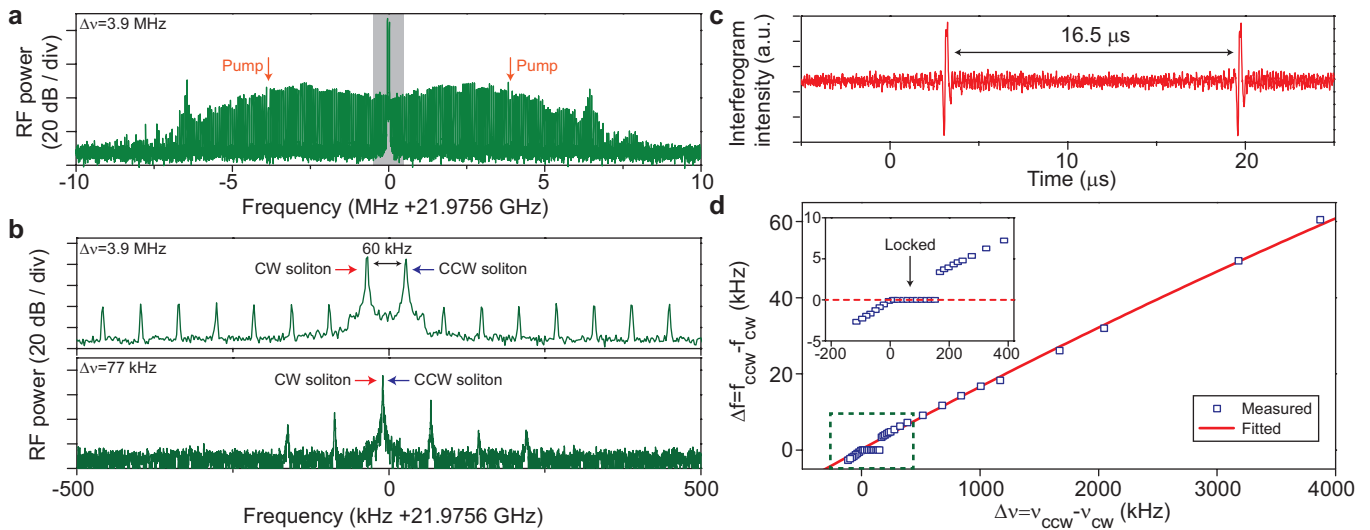}
\caption{{\bf Counter-propagating solitons with independently tuned repetition rates.} {\bf a,} Electrical spectrum of photo-detected CW and CCW soliton pulse streams with pump frequency difference $\Delta \nu = 3.9$ MHz. Strong central peaks give the repetition rate of each soliton. The weaker spectral lines occurring over a broader spectral range are inter-soliton beat frequencies. Beat frequencies produced by the pump line of one soliton beating with higher and lower frequency comb lines that neighbor the pump line of the other soliton are indicated by arrows. These spectral lines are shifted by $\Delta \nu = \pm$ 3.9MHz relative to the two, strong repetition-rate lines. {\bf b,} Upper trace is gray-banded region from fig. 2a. The pair of strong central peaks give the CW and CCW soliton repetition rates. Lower trace is the same electrical spectrum when the soliton repetition rates have locked to the same frequency. Pump frequencies differ by $\Delta \nu = $ 77 kHz. {\bf c,} Temporal interferogram of the baseband inter-soliton beat signal under unlocked condition in fig. 2a. {\bf d,} Plot of the difference in CW and CCW repetition rates versus versus difference in pump frequencies. The red line is a fit using the model in Methods. The inset shows that the two soliton repetition rates are locked over approximately 150 kHz pump difference frequency range.}
\label{figure2}
\end{figure*}

All whispering-gallery microresonators feature clockwise (CW) and counter-clockwise (CCW) optical whispering gallery modes, and this degree of freedom has not been explored for generation of soliton pulse trains. In this work counter-propagating (CP) solitons are generated by counter-pumping on a single microcavity resonance (fig.1a). Because DKSs are phase coherent with their respective optical pump, the tuning of two counter-propagating pumps causes an offset in the optical frequency of the two soliton pulse streams. Also, on account of the Raman-induced soliton self-frequency shift (SSFS), the repetition rate of a soliton pulse stream depends on the detuning of the pump frequency relative to the resonant frequency being pumped \cite{yang2016spatial,yi2016theory}. As a result, the pulse rate of each CP soliton pulse stream can be independently controlled. 

Besides independent repetition rate control there are two soliton phase locking effects that are observed. In both of these locked states, the soliton streams are also optically phase locked even though the soliton frequencies reside on a distinctly different grid of optical frequencies. In the first, the CW and CCW solitons are observed to phase lock with identical repetition rates. In the second locking effect the CP solitons experience relative rate locking at different repetition rates.  As a result, the microresonator produces two, soliton streams having different repetition rates but with high relative coherence. This form of locking is potentially useful in dual comb spectroscopy and in laser ranging systems \cite{coddington2009rapid} (LIDAR) where it would eliminate the need for independent and mutually-locked frequency combs.

The counter-propagating solitons are typically several-hundred femtoseconds in duration and the microcavity round-trip time is 46 ps. To produce the solitons, a continuous-wave fiber laser is amplified and split using a directional coupler so as to pump CW and CCW modes of a microcavity resonance using a fiber taper coupler (see experimental setup in fig. 1b). Two acousto-optic modulators (AOM) are used to control the pump power and frequencies in each pumping direction. The residual transmitted pump power is filtered by a fiber Bragg grating filter (FBG).

\begin{figure*}
\captionsetup{singlelinecheck=no, justification = RaggedRight}
\includegraphics[width=18cm]{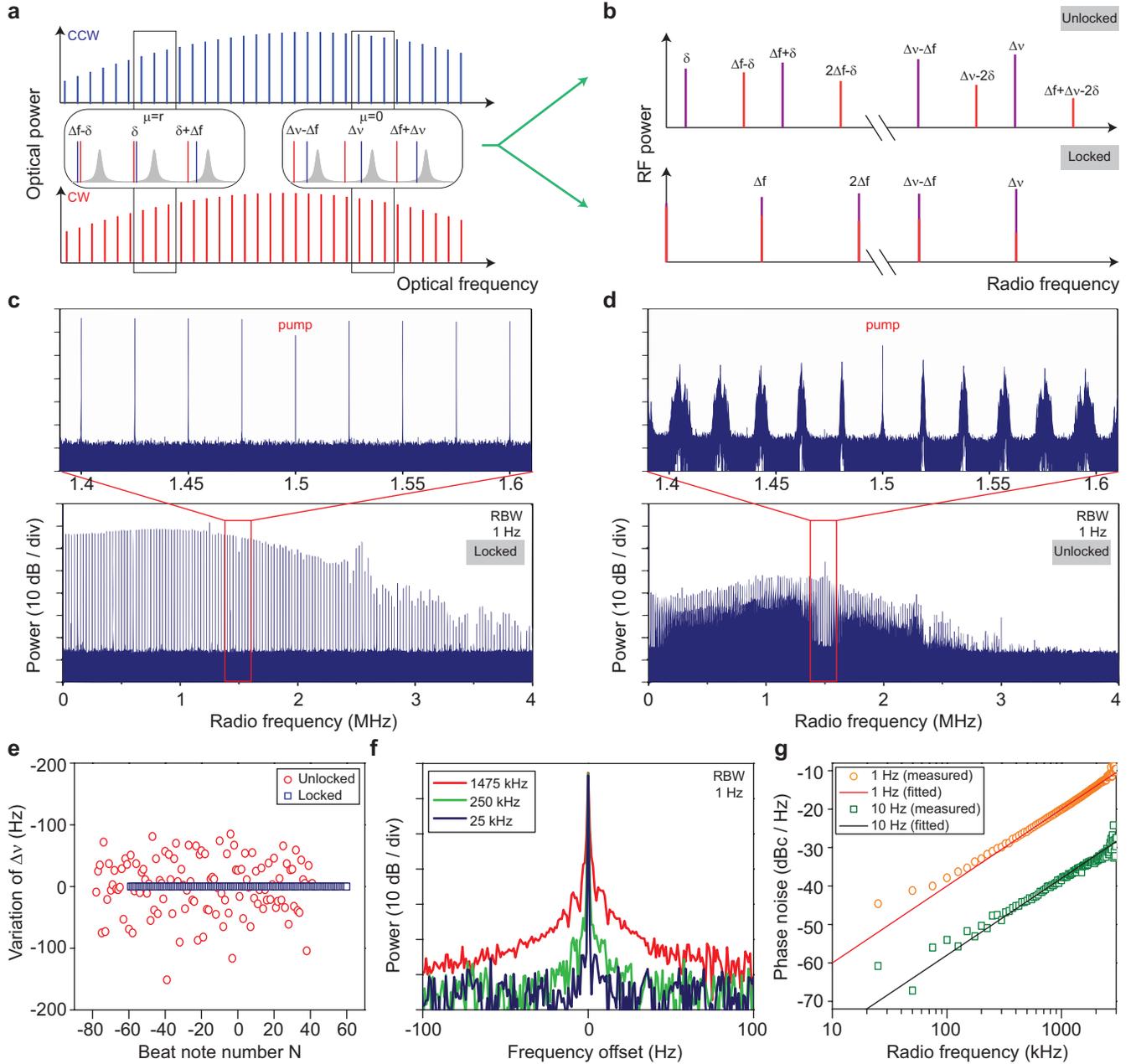}
\caption{{\bf Counter-propagating soliton phase locking at different repetition rates.} {\bf a,} Schematic view of the counter-propagating soliton comb lines. $\Delta \nu$ and $\Delta f$ denote the pump frequency and repetition rate differences, respectively, and $\mu$ is mode number relative to the pump mode ($\mu =0$). {\bf b,} Illustration of inter-soliton radio-frequency (RF) beatnotes produced under locked and unlocked conditions. {\bf c,} Measured RF beatnotes of locked CP solitons ($\Delta \nu = 1.5$MHz, $\Delta f=25$ kHz). {\bf d,} Measured RF beatnotes of unlocked CP solitons ($\Delta \nu$ = 1.5 MHz). {\bf e,} Measured beat-note spacing for locked and unlocked conditions plotted versus beatnote number. {\bf f,} High-resolution, zoom-in spectrum of RF beatnotes in {\bf c}. The corresponding beat note frequency is provided in the legend (25kHz is the fundamental beat note frequency). {\bf g,} Phase noise of the beatnotes at 1 Hz and 10 Hz offset frequencies in the phase noise spectrum plotted versus beatnote frequency. The fitting lines have an $f^2$ dependence.}
\label{figure3}
\end{figure*}

The CP solitons are stabilized indefinitely using the active capture technique \cite{yi2016active}. It is found that application of this locking technique to only one of the soliton pulse streams automatically locks the other pulse stream. In figure 1c the measured optical spectra and the autocorrelation traces (insets) for typical CW and CCW soliton streams are shown. The system can be controllably triggered and locked with a single or a specified number  of solitons in each propagation direction. The microresonator, a high-Q silica wedge design \cite{lee2012chemically} with 3 mm diameter, has anomalous dispersion at the pumping wavelength near 1.55 microns and is engineered to produce minimal avoided mode crossings over the optical band of the solitons \cite{yi2015soliton}. 

A feature of the dissipative Kerr soliton when viewed as a mode-locked frequency comb, is that the pump provides one of the comb frequencies and is therefore coherent with the soliton. Non-degenerate counter-pumping therefore introduces a controlled frequency offset between CW and CCW solitons. Because the counter-propagating pumps are derived from a single laser source, the mutual optical coherence of CW and CCW pump comb teeth is excellent and determined primarily by the stability of the radio-frequency signals used to drive the AOMs shown in fig. 1b. 

A key parameter that controls the soliton properties is the cavity-pump detuning frequency $\delta \omega_{\rm cw, ccw} = \omega_0 - \omega_{\rm cw, ccw}$ where $\omega_0$ is the cavity resonant frequency which is pumped and $\omega_{\rm cw, ccw}$ are the CW and CCW pump frequencies. Soliton pulse width \cite{herr2014temporal,yi2015soliton}, average power \cite{herr2014temporal,yi2015soliton}, and self-Raman-shift\cite{milian2015solitons,karpov2016raman,yi2016theory} depend upon this detuning. In cases where the self-Raman shift is strong, the soliton repetition rate also depends upon the cavity-pump detuning \cite{yi2016theory,yang2016spatial} and the CW and CCW soliton repetition rates ($f_\mathrm{cw}$ and $f_\mathrm{ccw}$) can be separately controlled by tuning of the respective pump frequencies.

To measure the CW and CCW soliton repetition rates, their pulse streams are combined and simultaneously photodetected. The electrical spectrum of the photocurrent is shown in fig. 2a when the difference in pumping frequencies is set to $\Delta \nu \equiv (\omega_{\mathrm{ccw}}-\omega_{\mathrm{cw}}) / 2 \pi $ = 3.9 MHz and $\delta \omega_\mathrm{cw} \sim$ 20 MHz. A zoom-in of the spectrum in the upper panel of fig. 2b shows that two strong central spectral peaks differ by 60 kHz. These peaks are the fundamental repetition rates associated with the CW and CCW soliton streams. The weaker, non-central beats appearing in fig. 2a and the upper panel in fig. 2b are inter-soliton beat frequencies between comb teeth belonging to different soliton combs. These beat frequencies are equally separated by the difference in the repetition rates (60 kHz). As an aside, the maxima at the extreme wings of the spectrum are caused by the mode crossing distortion in the comb spectra seen in fig. 1c near 1542 nm. 

An interferogram showing the electrical time trace of the co-detected dual-soliton pulse streams is shown in fig. 2c. This time trace can be understood as a stroboscopic interference of the respective soliton pulses on the detector. The strobing occurs at the rate difference ($\Delta f \equiv (f_{\mathrm{ccw}}-f_{\mathrm{cw}}) $) of the two soliton streams giving the repetitive signal a period of 16.5 $\mu$s. By varying the pump detuning, $\Delta \nu$, it is possible to observe tuning of the repetition rate difference, $\Delta f$, as shown in fig. 2d. A theoretical fit discussed in Methods is provided in the figure. Near $\Delta \nu = 0$ locking of the repetition rates is observed over a range of $\Delta \nu$ around 150 kHz. The associated electrical zoom-in spectrum under this locked condition is shown in the lower panel of fig. 2b. Importantly, nearly all of the weaker peaks that appear in the unlocked spectrum shown in fig. 2a disappear as a result of locking. This can be understood to result from the high relative temporal stability of the two pulse streams. In particular, under the locking condition, inter-soliton pulse mixing on the photo-detector, which is guaranteed under conditions of unequal repetition rates, now requires strict spatial-temporal alignment of the two pulse streams at the detector.  Consistent with this physical picture, the interferogram trace is observed to show no periodic strobing behavior. This locking behavior is believed to occur when pump light from a given pumping direction is backscattered into the opposing direction where the two pump signals can mix by the Kerr-effect.  This induces four-wave mixing sidebands on the soliton comb lines that subsequently induce locking. 

In addition to locking at identical repetition rates (degenerate locking), the soliton pulse streams are observed to lock when their repetition rates are different. Fig. 3a illustrates the principle of this locking mechanism. Therein, soliton spectra for CW and CCW directions are presented. A zoom-in of the higher frequency portion of the spectra is shown in which the respective soliton spectral lines are superimposed next to shaded areas representing the cavity resonances. The mode index $\mu = 0$, which is by convention the optical pump, is also indicated. As required for DKS generation, this pump frequency and the other soliton comb teeth are red-detuned in frequency relative to their respective cavity resonances.  

At $\mu = 0$, the two pump lines are separated by the pump frequency difference, $\Delta \nu$. Under conditions where these pump frequencies are well separated so that degenerate rate locking does not occur (see fig. 2d), the soliton having the more strongly red-detuned pump will feature a slightly lower repetition rate on account of the self-Raman-effect discussed above and in Methods. Accordingly, the CW and CCW comb lines will shift in frequency so as to become more closely spaced as $\mu$ decreases. For a certain negative value of $\mu$ the CW and CCW comb lines will achieve closest spectral separation. In the illustration, this occurs at comb tooth $\mu = r$ where CW and CCW comb lines have frequency separation $\delta = \Delta \nu + r \Delta f$. Backscattering within the resonator will couple power between these nearly resonant lines. This power coupling is shown in Methods to induce locking with a corresponding bandwidth.

Because the original comb teeth at $\mu = 0$ are derived from the same laser, the additional locking at $\mu=r$ causes the CW and CCW solitons to be mutually phase locked. Moreover, the difference in the soliton repetition rates must be an integer fraction (1/$|r|$) of pump frequency difference, 
\begin{equation}
\Delta f = -\Delta \nu / r,
\end{equation}
This result shows that pulse rates have a relative stability completely determined by the radio frequency signal used to set the pump frequency offset. Accordingly, the beat signal between the CCW and CW solitons exhibits very high stability when the system is locked in this way. The above relation also shows that the locked CP solitons play the role of a frequency divider of the pump frequency difference into the pulse-rate difference frequency. The phase noise of the rate difference is therefore $r^2$ lower than the phase noise of the relative pump signal,
\begin{equation}
S_{\Delta f}=\frac{1}{r^2}S_{\Delta \nu}.
\label{pn}
\end{equation}
where $S_{\Delta f}$ and $S_{\Delta\nu}$ are the phase noise spectral density functions of the inter-soliton fundamental beat signal and the pump difference signal. 

Fig. 3b illustrates the effect of the locking condition on the electrical spectrum produced by photodetection of combined CCW and CW soliton streams.  Under unlocked conditions, the electrical spectrum will feature two distinct spectra with spacing $\Delta f$. However, under locked conditions the difference in the frequency of the comb teeth at $\mu = 0$ (i.e., optical pumps) is an integer multiple of the difference in the repetition rates. As result, the two electrical spectra merge to form a single spectrum. Fig. 3c shows a typical measured RF spectrum in the locked state. It is obtained by Fourier transforming an interferogram recorded over 1s. A set of equidistant spectral lines is observed with a 50 dB signal-to-noise ratio (SNR) at the 1 Hz resolution bandwidth (RBW). In this measurement, $\Delta \nu$ is set to be 1.5 MHz which is 60 times $\Delta f =$ 25 kHz.  Meanwhile, the unlocked state shown in fig. 3d features relatively noisier spectral lines and lower SNR. This noise results from fluctuations of the absolute pump frequencies which induce fluctuations in the two Raman-shifted repetition rates \cite{yi2016single}. The resulting noise is multiplied with each comb tooth index relative to the pump comb tooth.  Fig. 3e plots the spacing between the RF comb lines. It shows collapse to a sub-Hz stability under the locked condition.

In strong contrast to the unlocked case, the spectral line beatnotes in fig. 3c actually improve in stability with decreasing order relative to the pump line. This a consequence of the frequency division noted in eq. (1) and eq. (2). In particular, the lowest frequency inter-soliton beatnote features the minimum linewidth as shown in fig. 3f. Furthermore, to confirm the scaling of phase noise with the frequency division given in eq. (2), fig. 3g shows the phase noise spectral density versus the spectral beatnote number measured at two phase-noise offset frequencies (1 Hz and 10 Hz). The dependence follows the predicted quadratic form typical of a frequency divider. It is noted that the  inferred linewidth for the lowest order beatnote is 40 $\mu$Hz ( assuming that it is limited by white frequency noise).

Counter-propagating solitons have been demonstrated in a high-Q optical microresonator. Both the repetition rates and the spectral location for the clockwise and counter-clockwise directions are independently tuned by tuning of the corresponding optical pumping frequencies. Two distinctly different locking phenomena have been observed while tuning the soliton repetition frequencies. In the first, the repetition rates lock to the same value. The pumping frequencies are different when this locking occurs so that the two soliton comb spectra are offset slightly in the optical frequency, but have identical comb line spacings. The interferogram of the two pulse trains has no baseband time dependence when this locking occurs. In the second form of locking, the pumps are typically tuned apart to larger difference frequencies and the solitons are observed to lock at different repetition rates with a difference that divides into the pump-frequency difference. The origin of this locking is associated with optical locking of two comb teeth, one from each soliton. Since the two pumps are derived from the same laser, this additional comb tooth locking effectively results in the two comb spectra being locked at two different positions in their spectra. The resulting high level of mutual soliton coherence is observable in the base-band inter-soliton beat spectra which features very narrow spectral lines spaced by the difference in the locked soliton repetition rates. In effect, this second form of locking creates two frequency combs in the same device with distinct repetition rates and optical frequencies, but that are optically locked. It is potentially useful in dual comb spectroscopy and dual comb LIDAR applications where it would obviate the need for two separate frequency combs and the associated inter-comb locking hardware. Finally, it is noted that while single clockwise and counter-clock-wise solitons have been generated, it is also possible to create states containing multiple solitons.

\medskip

\noindent\textbf{Methods}

\begin{footnotesize}
\noindent{\bf Repetition rate control of CP solitons.} The Raman SSFS, $\Omega_\mathrm{R}$, is dependent on the pump-cavity detuning, $\delta \omega$, by \cite{yi2016theory}
\begin{equation}
\Omega_\mathrm{R}=-\frac{32D_1^2 \tau_\mathrm{R}}{15\kappa D_2}\delta\omega^2
\end{equation}
where $\tau_\mathrm{R}$ is the Raman shock time, $\kappa$ is the cavity decay rate, $D_1$ ($D_2$) is the free-spectral-range (second-order dispersion) at mode $\mu = 0$ (the pumping mode). The soliton repetition rate, $f$, is coupled to the SSFS as
\begin{equation}
2\pi f=D_1+\frac{\Omega_\mathrm{R} D_2}{D_1}
\end{equation}
Therefore the interferogram between the counter-propagating solitons with cavity-pump detuning $\delta\omega_{\mathrm{cw}}$ and $\delta\omega_{\mathrm{ccw}}$ has a repetition rate difference
\begin{equation}
\begin{split}
f_\mathrm{ccw}-f_\mathrm{cw}=-\frac{16 D_1 \tau_\mathrm{R}}{15\pi \kappa}(\delta\omega_\mathrm{ccw}^2-\delta\omega_\mathrm{cw}^2) \\ = -\frac{16 D_1 \tau_\mathrm{R}}{15\pi \kappa} (2 \delta \omega_\mathrm{ccw} \Delta \omega - \Delta \omega^2)
\end{split}
\end{equation}
The second form of this equation uses $\Delta \omega = \omega_\mathrm{ccw} - \omega_\mathrm{cw} = 2 \pi \Delta \nu$ and is applied for the theoretical plot in fig. 2d.

\medskip

\noindent{\bf Locking of CP solitons}
The dissipative Kerr solitons are governed by the Lugiato-Lefever equation augmented by the Raman term \cite{lugiato1987spatial,karpov2016raman,yi2016theory}. The presence of scattering centers can induce coupling between the CP solitons as follows,
\begin{equation}
\begin{split}
\frac{\partial A(\phi, t)}{\partial t}= -(\frac{\kappa}{2} +i\delta \omega_\mathrm{A})A+i\frac{D_2}{2}\frac{\partial^2 A}{\partial {\phi}^2}+F
+ig|A|^2A  \\
+ig\tau_\mathrm{R}D_1A\frac{\partial |A|^2}{\partial \phi}+i\int_0^{2\pi} \Gamma(\theta)B(\phi-2\theta,t)e^{-i\Delta\omega t}\mathrm{d}\theta
\end{split}
\label{LLEA}
\end{equation}
\begin{equation}
\begin{split}
\frac{\partial B(\phi, t)}{\partial t}= -(\frac{\kappa}{2} +i\delta \omega_\mathrm{B})B+i\frac{D_2}{2}\frac{\partial^2 B}{\partial {\phi}^2}+F
+ig|B|^2B  \\
+ig\tau_\mathrm{R}D_1B\frac{\partial |B|^2}{\partial \phi}+i\int_0^{2\pi} \Gamma(\theta) A(\phi+2\theta,t)e^{i\Delta\omega t}\mathrm{d}\theta
\end{split}
\end{equation}
Here $A$ and $B$ denote the slowly varying field envelopes of the CW and CCW solitons, respectively. $\phi$ is the angular coordinate in the rotational frame \cite{herr2014temporal}. $g$ is the normalized Kerr nonlinear coefficient \cite{herr2014temporal,yi2016theory}, $F$ denotes the normalized continuous-wave pump term and $\Gamma(\theta)$ represents the backscattering coefficient in the lab frame $\theta$.

Considering the spectral misalignment of CP soliton comb lines presented in fig. 3a, it is assumed that only the $r$-th comb lines will induce inter-soliton coupling.  Accordingly, the equation of motion for the soliton field amplitude $A$, eq. \ref{LLEA}, is reduced to the following,
\begin{equation}
\begin{split}
\frac{\partial A(\phi, t)}{\partial t}= -(\frac{\kappa}{2} +i\delta \omega_\mathrm{A})A+i\frac{D_2}{2}\frac{\partial^2 A}{\partial {\phi}^2}+F
+ig|A|^2A  \\
+ig\tau_\mathrm{R}D_1A\frac{\partial |A|^2}{\partial \phi}+iGb_r e^{ir\phi}
\label{eq:LLEA}
\end{split}
\end{equation}
where the expansion $B(\phi,t)e^{i\Delta\omega t}=\sum_\mu b_\mu e^{i\mu\phi}$ is used to extract the $r$-th comb line from soliton field $B$. A similar equation of motion to eq.(\ref{eq:LLEA}) holds for the amplitude $B$ (with corresponding expansion $A(\phi,t)=\sum_\mu a_\mu e^{i\mu\phi}$). The coupling coefficient $G=\int \Gamma(\theta)\exp(-2ir\theta)\mathrm{d}\theta$.

The soliton field amplitude in the presence of the soliton self-frequency shift can be expressed as, \cite{matsko2013timing,yi2016theory}
\begin{equation}
A=B_s \mathrm{sech}[(\phi-\phi_\mathrm{Ac})/D_1\tau_s]e^{i\mu_\mathrm{A}(\phi-\phi_\mathrm{Ac})+i\psi_\mathrm{A}}
\end{equation}
where $B_s$ and $\tau_s$ are the pulse amplitude and duration, respectively. $\mu_A$ is the mode number of the soliton spectral maximum ($\mu=0$ is the mode number of the pump mode). This mode number is related to the soliton self-frequency shift by $\Omega_\mathrm{R} = \mu_\mathrm{A} D_1$.  $\psi_\mathrm{A}$ is a constant phase determined by the pump \cite{herr2014temporal,yi2016theory}. $\phi_\mathrm{{Ac}}$ is the peak position of the CW soliton, which is coupled to $\mu_\mathrm{A}$ by \cite{yi2016theory}
\begin{equation}
\frac{\partial\phi_\mathrm{Ac}}{\partial t}=\mu_\mathrm{A}D_2.
\label{eq:repratecouple}
\end{equation}

The soliton energy $E_\mathrm{A}$ and the spectral maximum mode number $\mu_\mathrm{A}$ are given by
\begin{equation}
E_\mathrm{A} = \sum_\mu|a_\mu|^2 = \frac{1}{2\pi}\int_{-\pi}^{+\pi}|A|^2\mathrm{d}\phi = B_s^2\tau_sD_1/\pi
\label{eq:E}
\end{equation}
\begin{equation}
\mu_\mathrm{A} = \frac{\sum_\mu\mu|a_\mu|^2}{E_\mathrm{A}} = \frac{-i}{4\pi E_\mathrm{A}}\int_{-\pi}^{+\pi}(A^*\frac{\partial A}{\partial \phi}-A\frac{\partial A^*}{\partial \phi})\mathrm{d}\phi
\label{eq:muc}
\end{equation} 
Taking the time derivative of eq.(\ref{eq:muc}) and substituting $\partial A/ \partial t$ using eq.(\ref{eq:LLEA}), the equation of motion for $\mu_\mathrm{A}$ is obtained as
\begin{equation}
\begin{split}
\frac{\partial \mu_\mathrm{A}}{\partial t} &= -\kappa \mu_\mathrm{A} -\frac{g\tau_\mathrm{R}D_1}{2\pi E_\mathrm{A}}\int_{-\pi}^{+\pi}(\frac{\partial |A|^2}{\partial \phi})^2 \mathrm{d} \phi \\
&-\frac{1}{2\pi E_\mathrm{A}} \int_{-\pi}^{+\pi} (G^* b_r^* e^{-ir\phi}\frac{\partial A}{\partial \phi}-irGA^*b_r e^{ir\phi})\mathrm{d}\phi
\label{eq:pmuc}
\end{split}
\end{equation}
The second term on the right-hand-side corresponds to the steady-state Raman-induced center shift \cite{karpov2016raman,yi2016theory} and is denoted by $\kappa R_\mathrm{A}$. The third term is the soliton spectral shift caused by coupling to the opposing CP soliton through its comb tooth $b_r$. By using $A=\sum_\mu a_\mu e^{i\mu\phi}$, eq. (\ref{eq:pmuc}) yields,
\begin{equation}
\begin{split}
\frac{\partial\mu_\mathrm{A}}{\partial t}&=-\kappa\mu_\mathrm{A}+\kappa R_\mathrm{A}-\frac{i r}{E_\mathrm{A}}(a_r b_r^*G^*-\mathrm{c.c.})\\
&=-\kappa\mu_\mathrm{A}+\kappa R_\mathrm{A}+\frac{2 r}{E_\mathrm{A}}|a_r b_r G|\sin\Theta.
\end{split}
\label{eq:muAmotion}
\end{equation}
where $\Theta=(\psi_{r\mathrm{A}}-\psi_{r\mathrm{B}}-\psi_G)$ with the phases, $\psi_\mathrm{rA}$ and $\psi_\mathrm{rB}$, of the comb lines $a_r$ and $b_r$ given by the following expression,
\begin{equation}
\psi_\mathrm{rA}=\psi_\mathrm{A}-r\phi_\mathrm{Ac}.
\end{equation}
\begin{equation}
\psi_\mathrm{rB}=\psi_\mathrm{B}-r\phi_\mathrm{Bc} +\Delta \omega t.
\end{equation}
Also, $\Psi_G$ is the phase of the backscatter coefficient $G$. The time dependence of $\psi_\mathrm{rA}$ can be derived from eq. \ref{eq:repratecouple} as,
\begin{equation}
\frac{\partial\psi_\mathrm{rA}}{\partial t}=-r\frac{\partial\phi_\mathrm{Ac}}{\partial t}=-r\mu_\mathrm{A}D_2.
\end{equation}
Similarly, the derivative of the phase of $b_r$ is given by,
\begin{equation}
\frac{\partial\psi_\mathrm{rB}}{\partial t}=-r\mu_\mathrm{B}D_2+\Delta\omega.
\end{equation}
Therefore the time derivative of the phase term $\Theta=(\psi_{r\mathrm{A}}-\psi_{r\mathrm{B}}-\psi_G)$ is given by,
\begin{equation}
\frac{\partial\Theta}{\partial t}=\Delta\omega+rD_2(\mu_\mathrm{B}-\mu_\mathrm{A}) = 2 \pi  (\Delta\nu+ r \Delta f ) = 2 \pi \delta.
\label{eq:theta}
\end{equation}
Similar to eq.(\ref{eq:muAmotion}), a parallel equation exists for the soliton $B$ and is given by, 
\begin{equation}
\begin{split}
\frac{\partial\mu_\mathrm{B}}{\partial t}=-\kappa\mu_\mathrm{B}+\kappa R_\mathrm{B}-\frac{2 r}{E_\mathrm{B}}|a_r b_r G|\sin\Theta.
\end{split}
\label{eq:muBmotion}
\end{equation}
Taking a time derivative of eq.(\ref{eq:theta}) and using eq.(\ref{eq:muAmotion}) and eq.(\ref{eq:muBmotion}) gives the following equation of motion for the relative phase $\Theta$,
\begin{equation}
\frac{\partial ^2 \Theta }{\partial t^2} + \kappa \frac{\partial \Theta }{\partial t} = -2r^2 D_2 (\frac{1}{E_\mathrm{A}}+\frac{1}{E_\mathrm{B}})|a_r b_r G|\sin\Theta + 2\pi\kappa \delta',
\label{eq:thetamotion}
\end{equation}
where $2\pi\delta' =\Delta\omega + rD_2(R_\mathrm{B}-R_\mathrm{A})$ is the frequency difference between the $r_\mathrm{th}$ comb lines induced by the shifted pumps and Raman SSFS when the CP solitons have no interaction. The above equation is similar to the Alder equation of injection locking \cite{adler1946study}, only with an additional second order time-derivative term. Setting the time derivatives of $\Theta$ equal to zero gives the locking bandwidth, $\omega_L$, of $\delta'$ as
\begin{equation}
\omega_L=4\pi|\delta'_{\mathrm{max}}|=\frac{4r^2D_2}{\kappa}(\frac{1}{E_\mathrm{A}}+\frac{1}{E_\mathrm{B}})|a_r b_r G|.
\end{equation}
Moreoever, eq.(\ref{eq:theta}) gives $\delta = 0$ so that the pump frequency difference $\Delta \nu$ is divided by the repetition rate difference as follows,
\begin{equation}
\Delta f=-\frac{\Delta \nu}{r},
\end{equation}
{\it which is eq. (1) in the main text.}

\medskip

{\noindent \bf Parameters.} In the measurement, the loss rate is $\kappa/2\pi=1.5$ MHz. $D_2/2\pi=16$ kHz and $r=-60$. For a soliton with $\tau_s=150$ fs, the mode number of the Raman SSFS is $\mu_\mathrm{R}\sim -20$ and the ratio $|a_r|^2/E_\mathrm{A}=D_1\tau_s\mathrm{sech}^2[\pi (r-\mu_\mathrm{R}) D_1\tau_s/2]/8\sim 7\times 10^{-4}$. As the CP solitons have similar powers, the locking bandwidth is estimated as $\omega_L\sim |G|/4$. In this case a backscattering rate of 4 kHz can provide a 1 kHz locking bandwidth.

\end{footnotesize}

\medskip

\noindent\textbf{Acknowledgment}

\noindent The authors gratefully acknowledge the Defense Advanced Research Projects Agency under the PULSE and DODOS programs, NASA, the Kavli Nanoscience Institute.

\bibliography{ref}

\begin{thebibliography}{10}
\expandafter\ifx\csname url\endcsname\relax
  \def\url#1{\texttt{#1}}\fi
\expandafter\ifx\csname urlprefix\endcsname\relax\def\urlprefix{URL }\fi
\providecommand{\bibinfo}[2]{#2}
\providecommand{\eprint}[2][]{\url{#2}}

\bibitem{herr2014temporal}
\bibinfo{author}{Herr, T.} \emph{et~al.}
\newblock \bibinfo{title}{Temporal solitons in optical microresonators}.
\newblock \emph{\bibinfo{journal}{Nat. Photon.}} \textbf{\bibinfo{volume}{8}},
  \bibinfo{pages}{145--152} (\bibinfo{year}{2014}).

\bibitem{yi2015soliton}
\bibinfo{author}{Yi, X.}, \bibinfo{author}{Yang, Q.-F.}, \bibinfo{author}{Yang,
  K.~Y.}, \bibinfo{author}{Suh, M.-G.} \& \bibinfo{author}{Vahala, K.}
\newblock \bibinfo{title}{Soliton frequency comb at microwave rates in a high-q
  silica microresonator}.
\newblock \emph{\bibinfo{journal}{Optica}} \textbf{\bibinfo{volume}{2}},
  \bibinfo{pages}{1078--1085} (\bibinfo{year}{2015}).

\bibitem{brasch2016photonic}
\bibinfo{author}{Brasch, V.} \emph{et~al.}
\newblock \bibinfo{title}{Photonic chip--based optical frequency comb using
  soliton cherenkov radiation}.
\newblock \emph{\bibinfo{journal}{Science}} \textbf{\bibinfo{volume}{351}},
  \bibinfo{pages}{357--360} (\bibinfo{year}{2016}).

\bibitem{wang2016intracavity}
\bibinfo{author}{Wang, P.-H.} \emph{et~al.}
\newblock \bibinfo{title}{Intracavity characterization of micro-comb generation
  in the single-soliton regime}.
\newblock \emph{\bibinfo{journal}{Opt. Express}} \textbf{\bibinfo{volume}{24}},
  \bibinfo{pages}{10890--10897} (\bibinfo{year}{2016}).

\bibitem{joshi2016thermally}
\bibinfo{author}{Joshi, C.} \emph{et~al.}
\newblock \bibinfo{title}{Thermally controlled comb generation and soliton
  modelocking in microresonators}.
\newblock \emph{\bibinfo{journal}{Opt. Lett.}} \textbf{\bibinfo{volume}{41}},
  \bibinfo{pages}{2565--2568} (\bibinfo{year}{2016}).

\bibitem{ankiewicz2008dissipative}
\bibinfo{author}{Ankiewicz, A.} \& \bibinfo{author}{Akhmediev, N.}
\newblock \emph{\bibinfo{title}{Dissipative Solitons: From Optics to Biology
  and Medicine}} (\bibinfo{publisher}{Springer}, \bibinfo{year}{2008}).

\bibitem{leo2010temporal}
\bibinfo{author}{Leo, F.} \emph{et~al.}
\newblock \bibinfo{title}{Temporal cavity solitons in one-dimensional kerr
  media as bits in an all-optical buffer}.
\newblock \emph{\bibinfo{journal}{Nat. Photon.}} \textbf{\bibinfo{volume}{4}},
  \bibinfo{pages}{471--476} (\bibinfo{year}{2010}).

\bibitem{milian2015solitons}
\bibinfo{author}{Mili{\'a}n, C.}, \bibinfo{author}{Gorbach, A.~V.},
  \bibinfo{author}{Taki, M.}, \bibinfo{author}{Yulin, A.~V.} \&
  \bibinfo{author}{Skryabin, D.~V.}
\newblock \bibinfo{title}{Solitons and frequency combs in silica microring
  resonators: Interplay of the raman and higher-order dispersion effects}.
\newblock \emph{\bibinfo{journal}{Phys. Rev. A}} \textbf{\bibinfo{volume}{92}},
  \bibinfo{pages}{033851} (\bibinfo{year}{2015}).

\bibitem{karpov2016raman}
\bibinfo{author}{Karpov, M.} \emph{et~al.}
\newblock \bibinfo{title}{Raman self-frequency shift of dissipative kerr
  solitons in an optical microresonator}.
\newblock \emph{\bibinfo{journal}{Phys. Rev. Lett.}}
  \textbf{\bibinfo{volume}{116}}, \bibinfo{pages}{103902}
  (\bibinfo{year}{2016}).

\bibitem{yi2016theory}
\bibinfo{author}{Yi, X.}, \bibinfo{author}{Yang, Q.-F.}, \bibinfo{author}{Yang,
  K.~Y.} \& \bibinfo{author}{Vahala, K.}
\newblock \bibinfo{title}{Theory and measurement of the soliton self-frequency
  shift and efficiency in optical microcavities}.
\newblock \emph{\bibinfo{journal}{Opt. Lett.}} \textbf{\bibinfo{volume}{41}},
  \bibinfo{pages}{3419--3422} (\bibinfo{year}{2016}).

\bibitem{yang2016stokes}
\bibinfo{author}{Yang, Q.-F.}, \bibinfo{author}{Yi, X.}, \bibinfo{author}{Yang,
  K.~Y.} \& \bibinfo{author}{Vahala, K.}
\newblock \bibinfo{title}{Stokes solitons in optical microcavities}.
\newblock \emph{\bibinfo{journal}{Nat. Phys.}} \textbf{\bibinfo{volume}{13}},
  \bibinfo{pages}{53--57} (\bibinfo{year}{2017}).

\bibitem{matsko2016optical}
\bibinfo{author}{Matsko, A.~B.}, \bibinfo{author}{Liang, W.},
  \bibinfo{author}{Savchenkov, A.~A.}, \bibinfo{author}{Eliyahu, D.} \&
  \bibinfo{author}{Maleki, L.}
\newblock \bibinfo{title}{Optical cherenkov radiation in overmoded
  microresonators}.
\newblock \emph{\bibinfo{journal}{Opt. Lett.}} \textbf{\bibinfo{volume}{41}},
  \bibinfo{pages}{2907--2910} (\bibinfo{year}{2016}).

\bibitem{yang2016spatial}
\bibinfo{author}{Yang, Q.-F.}, \bibinfo{author}{Yi, X.}, \bibinfo{author}{Yang,
  K.~Y.} \& \bibinfo{author}{Vahala, K.}
\newblock \bibinfo{title}{Spatial-mode-interaction-induced dispersive-waves and
  their active tuning in microresonators}.
\newblock \emph{\bibinfo{journal}{Optica}} \textbf{\bibinfo{volume}{3}},
  \bibinfo{pages}{1132--1135} (\bibinfo{year}{2016}).

\bibitem{cole2016soliton}
\bibinfo{author}{Cole, D.~C.}, \bibinfo{author}{Lamb, E.~S.},
  \bibinfo{author}{Del'Haye, P.}, \bibinfo{author}{Diddams, S.~A.} \&
  \bibinfo{author}{Papp, S.~B.}
\newblock \bibinfo{title}{Soliton crystals in kerr resonators}.
\newblock \emph{\bibinfo{journal}{arXiv preprint arXiv:1610.00080}}
  (\bibinfo{year}{2016}).

\bibitem{kippenberg2011microresonator}
\bibinfo{author}{Kippenberg, T.~J.}, \bibinfo{author}{Holzwarth, R.} \&
  \bibinfo{author}{Diddams, S.}
\newblock \bibinfo{title}{Microresonator-based optical frequency combs}.
\newblock \emph{\bibinfo{journal}{Science}} \textbf{\bibinfo{volume}{332}},
  \bibinfo{pages}{555--559} (\bibinfo{year}{2011}).

\bibitem{brasch2017self}
\bibinfo{author}{Brasch, V.}, \bibinfo{author}{Lucas, E.},
  \bibinfo{author}{Jost, J.~D.}, \bibinfo{author}{Geiselmann, M.} \&
  \bibinfo{author}{Kippenberg, T.~J.}
\newblock \bibinfo{title}{Self-referenced photonic chip soliton kerr frequency
  comb}.
\newblock \emph{\bibinfo{journal}{Light Sci Appl.}}
  \textbf{\bibinfo{volume}{6}}, \bibinfo{pages}{e16202} (\bibinfo{year}{2017}).

\bibitem{liang2015high}
\bibinfo{author}{Liang, W.} \emph{et~al.}
\newblock \bibinfo{title}{High spectral purity kerr frequency comb radio
  frequency photonic oscillator}.
\newblock \emph{\bibinfo{journal}{Nat. Commun.}} \textbf{\bibinfo{volume}{6}},
  \bibinfo{pages}{7957} (\bibinfo{year}{2015}).

\bibitem{yi2016single}
\bibinfo{author}{Yi, X.}, \bibinfo{author}{Yang, Q.-F.},
  \bibinfo{author}{Zhang, X.}, \bibinfo{author}{Yang, K.~Y.} \&
  \bibinfo{author}{Vahala, K.}
\newblock \bibinfo{title}{Single-mode dispersive waves and soliton microcomb
  dynamics}.
\newblock \emph{\bibinfo{journal}{arXiv preprint arXiv:1610.08145}}
  (\bibinfo{year}{2016}).

\bibitem{suh2016microresonator}
\bibinfo{author}{Suh, M.-G.}, \bibinfo{author}{Yang, Q.-F.},
  \bibinfo{author}{Yang, K.~Y.}, \bibinfo{author}{Yi, X.} \&
  \bibinfo{author}{Vahala, K.~J.}
\newblock \bibinfo{title}{Microresonator soliton dual-comb spectroscopy}.
\newblock \emph{\bibinfo{journal}{Science}} \textbf{\bibinfo{volume}{354}},
  \bibinfo{pages}{600--603} (\bibinfo{year}{2016}).

\bibitem{dutt2016chip}
\bibinfo{author}{Dutt, A.} \emph{et~al.}
\newblock \bibinfo{title}{On-chip dual comb source for spectroscopy}.
\newblock \emph{\bibinfo{journal}{arXiv preprint arXiv:1611.07673}}
  (\bibinfo{year}{2016}).

\bibitem{pavlov2017soliton}
\bibinfo{author}{Pavlov, N.} \emph{et~al.}
\newblock \bibinfo{title}{Soliton dual frequency combs in crystalline
  microresonators}.
\newblock \emph{\bibinfo{journal}{Opt. Lett.}} \textbf{\bibinfo{volume}{42}},
  \bibinfo{pages}{514--517} (\bibinfo{year}{2017}).

\bibitem{yi2016active}
\bibinfo{author}{Yi, X.}, \bibinfo{author}{Yang, Q.-F.}, \bibinfo{author}{Youl,
  K.} \& \bibinfo{author}{Vahala, K.}
\newblock \bibinfo{title}{Active capture and stabilization of temporal solitons
  in microresonators}.
\newblock \emph{\bibinfo{journal}{Opt. Lett.}} \textbf{\bibinfo{volume}{41}},
  \bibinfo{pages}{2037--2040} (\bibinfo{year}{2016}).

\bibitem{coddington2009rapid}
\bibinfo{author}{Coddington, I.}, \bibinfo{author}{Swann, W.},
  \bibinfo{author}{Nenadovic, L.} \& \bibinfo{author}{Newbury, N.}
\newblock \bibinfo{title}{Rapid and precise absolute distance measurements at
  long range}.
\newblock \emph{\bibinfo{journal}{Nat. Photon.}} \textbf{\bibinfo{volume}{3}},
  \bibinfo{pages}{351--356} (\bibinfo{year}{2009}).

\bibitem{lee2012chemically}
\bibinfo{author}{Lee, H.} \emph{et~al.}
\newblock \bibinfo{title}{Chemically etched ultrahigh-q wedge-resonator on a
  silicon chip}.
\newblock \emph{\bibinfo{journal}{Nat. Photon.}} \textbf{\bibinfo{volume}{6}},
  \bibinfo{pages}{369--373} (\bibinfo{year}{2012}).

\bibitem{lugiato1987spatial}
\bibinfo{author}{Lugiato, L.~A.} \& \bibinfo{author}{Lefever, R.}
\newblock \bibinfo{title}{Spatial dissipative structures in passive optical
  systems}.
\newblock \emph{\bibinfo{journal}{Phys. Rev. Lett.}}
  \textbf{\bibinfo{volume}{58}}, \bibinfo{pages}{2209} (\bibinfo{year}{1987}).

\bibitem{matsko2013timing}
\bibinfo{author}{Matsko, A.~B.} \& \bibinfo{author}{Maleki, L.}
\newblock \bibinfo{title}{On timing jitter of mode locked kerr frequency
  combs}.
\newblock \emph{\bibinfo{journal}{Opt. Express}} \textbf{\bibinfo{volume}{21}},
  \bibinfo{pages}{28862--28876} (\bibinfo{year}{2013}).

\bibitem{adler1946study}
\bibinfo{author}{Adler, R.}
\newblock \bibinfo{title}{A study of locking phenomena in oscillators}.
\newblock \emph{\bibinfo{journal}{Proc. IEEE}} \textbf{\bibinfo{volume}{34}},
  \bibinfo{pages}{351--357} (\bibinfo{year}{1946}).

\end{thebibliography}
\end{document}